\documentstyle[12pt,twoside,fleqn,espcrc1,epsf]{article}

\newcommand{\AmS}{{\protect\the\textfont2
  A\kern-.1667em\lower.5ex\hbox{M}\kern-.125emS}}

\title{Nuclear dependence in transverse momentum distribution 
for Drell-Yan pair}

\author{Xiaofeng Guo\address{Department of Physics, Columbia 
           University; New York, NY 10027, USA}%
        \thanks{This work was supported by the U.S. Department of 
  Energy under contract No. DE-FG02-93ER40764.}}

\begin{document}
\maketitle

\begin{abstract}
In terms of multiple scattering picture, I compute  
the nuclear dependence in Drell-Yan  transverse
momentum distribution, $d\sigma / dQ^2 dq_T^2$, in hadron-nucleus 
collisions.  I  present the results for large $q_T$ region and
discuss the possible suppression in small $q_T$ region.
\end{abstract}

\section{INTRODUCTION}

Recently, it was found experimentally that 
the transverse momentum  distribution of Drell-Yan pair 
in hadron-nucleus collisions shows nuclear dependence 
not linearly proportional to $A$, the atomic number of the 
target \cite{NA10}. In addition, the average of squared 
transverse momentum, $\langle q_T^2 \rangle $, 
grows approximately as $A^{1/3}$ \cite{NA10,E772}. 
This observed non-linear nuclear dependence indicates that 
multiple-scattering is important in hadron-nucleus collisions.
In this talk, I present a calculation of double-scattering 
in QCD perturbation theory for 
nuclear dependence of Drell-Yan $q_T^2$ spectrum,
$d\sigma/dQ^2\,dq_T^2$.

Although double-scattering is in principle a power correction 
(or known as a high twist effect) to the leading single scattering, 
it is not necessary small because of $A^{1/3}$ enhancement 
from large nuclear size.
In terms of QCD factorization generalized to higher twist \cite{QS},
Luo, Qiu and Sterman (LQS) developed a consistent perturbative 
treatment of multiple scattering at parton level \cite{LQS}.
Since Drell-Yan pair does not interact strongly once produced, 
the observed non-linear nuclear dependence is a result of 
multiple scattering between the incoming beam parton and nuclear 
matter before the pair was produced.  
Drell-Yan $d\sigma/dQ^2\,dq_T^2$ has two observed physical scales:
$Q^2$ and $q_T^2$. When $Q^2$ and $q_T^2$ are both large,
$d\sigma/dQ^2\,dq_T^2$ is effectively an one-scale process.
Method developed by LQS \cite{LQS} for 
double scattering can be naturally applied.  However,
when $q_T^2 \ll Q^2$, a resummation of large 
$\ell n(Q^2/q_T^2)$ is necessary for a reliable prediction of
Drell-Yan $q_T^2$ spectrum \cite{CSS}.  

\section{PARTON LEVEL DOUBLE SCATTERING}         
\label{sec:dis}

\begin{figure}[htb]
\begin{minipage}[t]{80mm}
\epsfxsize=2in
\epsfbox{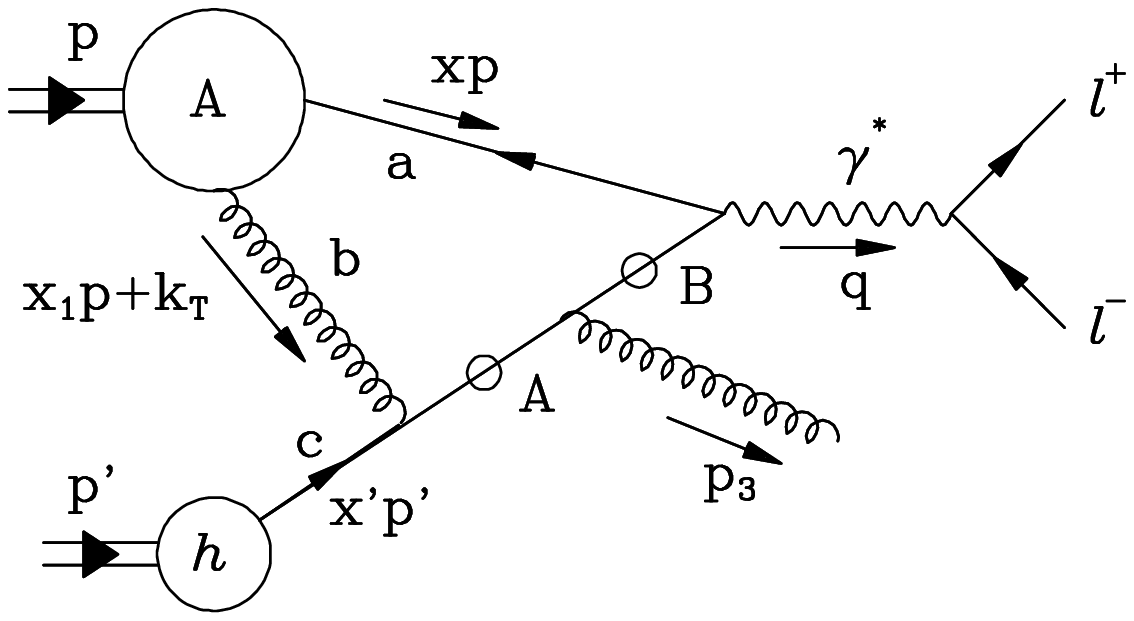}
\caption{A sample diagram for double-scattering amplitude.}
\label{fig1}
\end{minipage}
\hspace{\fill}
\begin{minipage}[t]{75mm}
\epsfxsize=2.0in
\epsfbox{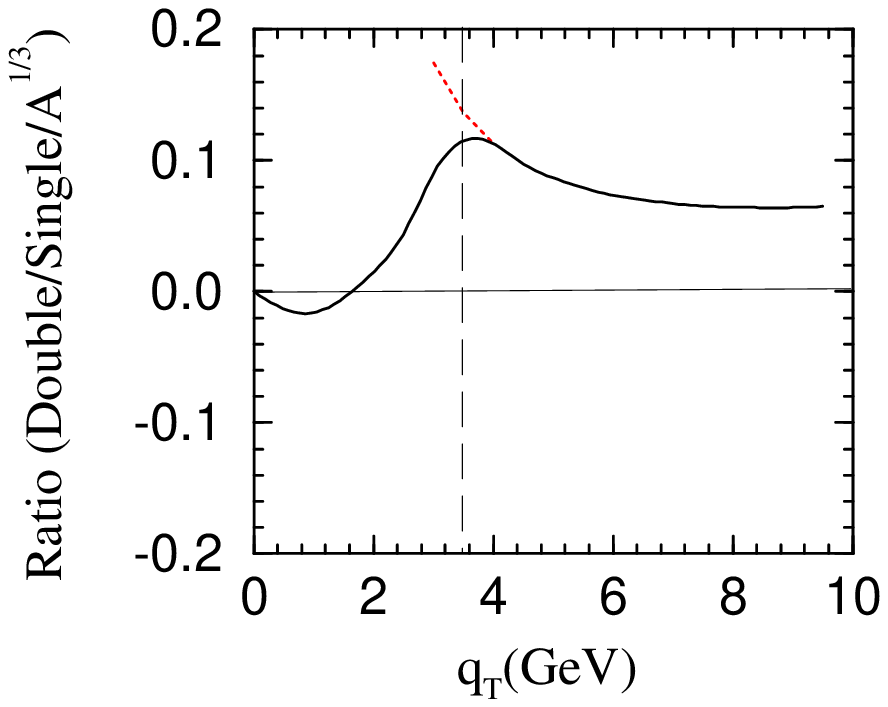}
\caption{Ratio of double-scattering to single-scattering.}
\label{fig2}
\end{minipage}
\end{figure}

Consider the Drell-Yan process in hadron-nucleus collisions,
$h(p')+A(p_A)\rightarrow l\bar{l}(q) +X$, where the lepton pair
has invariant mass ($Q$) and transverse momentum ($q_T$).
As an example, the scattering amplitude for the lowest 
order double-scattering, as shown in Fig.~\ref{fig1}, 
has the following general form: 
\begin{equation}
M \sim \int dx_1 \, \frac{1}{x_1-x_{1A}+i\epsilon} \,
\frac{1}{x_1-x_{1B}+i\epsilon} \, F(x_1,x), 
\label{sec5-e1}  
\end{equation}
where $x_1$ and $x$ are parton momentum fractions, 
as labeled in Fig.~\ref{fig1}, and $x_1$ needs to be integrated.  
The sum of $x_1$ and $x$ is fixed by the kinematics, 
such as $x'p'$, $p_3$ and $q$. 
In Eq.~(\ref{sec5-e1}), the function $F(x_1,x)$ is a non-vanishing 
and smooth function when $x_1=x_{1A}$ and/or $x_1=x_{1B}$, and it is
proportional to the parton fields of momentum $x_1p$ and $xp$
with $p\equiv p_A/A$.  Taking the pole at $x_1=x_{1A}$ 
(or at $x_1=x_{1B}$) 
corresponds to putting the propagator labeled by ``A'' (or ``B'') 
in Fig.~\ref{fig1} on its mass shell.
In the region of interests ($x's-2p_3\cdot p > 0$), both potential 
poles in Eq.~(\ref{sec5-e1}) are in the same half of the 
complex plane.  When $q_T \neq 0$, $x_{1A} \neq x_{1B}$. 
Contour integration of $dx_1$ yields 
\begin{equation}
M \, \sim \, \frac{F(x_{1A},x_{tot}-x_{1A})}{x_{1A}-x_{1B}} -
\frac{F(x_{1B},x_{tot}-x_{1B})}{x_{1A}-x_{1B}}
\quad \sim \, M_{soft-hard} -M_{double-hard} \ ,
\label{sec5-e3}
\end{equation}
where $x_{tot}$ is the sum of the total momentum fraction from the 
target, and is a function of $x'$, $p_3$ and $q$. 
Note that two amplitudes in Eq.~(\ref{sec5-e3})
have the opposite sign.

Amplitude $M_{soft-hard}$ corresponds to the residue at 
the pole ``A", at which $x_1 \approx 0$ and  $x \approx x_{tot}$.  
We call this type of double-scattering a soft-hard 
scattering.  In this case, 
the first scattering is effectively soft and 
not localized.  It represents a long range
correlation of the color field inside the nucleus. 
The second scattering, on the other hand, is
localized in a distance space $\sim 1/x_{tot}p \sim 1/Q$.
Because of the long range correlation of color fields,
the soft-hard double-scattering does not have the classical
double-scattering picture. Its contribution to the cross section can not 
be expressed in terms
of a product of two localized partonic cross sections. 

Amplitude $M_{double-hard}$ in Eq.~(\ref{sec5-e3}) 
corresponds to the residue at 
the pole ``B".  When $q_T \neq 0$, 
both $x_1=x_{1B}$ and $x=x_{tot}-x_{1B}$ are finite. We call this type of
double-scattering a double-hard scattering.  
In this case, both partonic scatterings are hard,
and they are localized at a distance $\sim 1/x_{1B}p$ and 
$\sim 1/(x_{tot}-x_{1B})p$, respectively.  Such a
double-hard subprocess resembles the classical double-scattering 
picture.  Its contribution to the cross section 
can be expressed in terms of a 
product of two localized partonic cross sections. 
 
From Eq.~(\ref{sec5-e3}), we see that the double-scattering 
contribution to the cross section has three types of  contributions: 
the soft-hard $\sim M^2_{soft-hard}$, the 
double-hard $\sim M^2_{double-hard}$, and the 
contribution from the interference terms. 
The interference contribution has an opposite sign in comparison with 
the other two contributions.
Consequently, the interference terms give nuclear {\it suppression}, 
while both the soft-hard and double-hard terms give the
nuclear {\it enhancement}.  From Eq.~(\ref{sec5-e3}), 
the interference contribution is proportional to 
$F^{*}(0,x_{tot})F(x_{1B},x_{tot}-x_{1B})$ 
plus its complex conjugate.
When $q_T$ is large, $x_{1B} \neq 0$, overlap in phase space 
for $F^{*}$ and $F$ is 
clearly small because of the difference in parton momenta. 
However, when $q_T\rightarrow 0$, $x_{1B} 
\sim O(q_T^2/Q^2)\rightarrow 0$. 
The overlap becomes large and the interference term becomes more important. 

\section{ENHANCEMENT IN LARGE $q_T$ REGION}

In large $q_T$ region, the interference between
the soft-hard and the double-hard subprocesses is not important, and 
consequently, the double-scattering contribution gives nuclear 
{\it enhancement} 
to the Drell-Yan transverse momentum distribution. 
The contribution from a double-hard
subprocess can be expresses as \cite{GUO}:
\begin{eqnarray}
\frac{d\sigma^{(DH)}}{dQ^2 dq_T^2 dy}
&=& \sum_{a,b,c=q,g,\bar{q}}\, \int dx' f_{c/h}(x') 
T^{DH}_{ab}(x_a, x_b, A) \cdot \frac{1}{2x'x_bs} \cdot
H^{DH}_{bc}(x_b,x')    \nonumber \\
&\ & \hspace{1.0in} \times
 \frac{12\pi\alpha_s^2}{x's+u-Q^2}\,\frac{1}{Q^2-u}\cdot 
\left(\frac{4\pi\alpha_{em}^2}{9Q^2} e_a^2 \right) \ .
\label{dh} 
\end{eqnarray}
In Eq.~(\ref{dh}), $f_{c/h}(x')$ is
the parton distribution of parton 
``c'' from the beam hadron $h$, and $x_a$, $x_b$ are the 
momentum fractions of the partons ``a'' and ``b'' 
from the target respectively. The $s, t, u$ are Mandelstam 
variables defined as
\begin{eqnarray}
s= (p+p')^2 =2p\cdot p'\, ; &t=(p'-q)^2 =-2p' \cdot q+Q^2\, ;& 
u = (p-q)^2 =-2p\cdot q+Q^2\ .
\end{eqnarray}
In Eq.~(\ref{dh}), 
$H_{bc}^{DH}(x_b,x')$ represents the calculated partonic part.  
Function $T^{DH}_{ab}(x_a, x_b, A)$ is a 
four parton matrix element (or correlation function) \cite{GUO}.
For a cold nucleus, we use the following model \cite{MQ}:
\begin{equation}
 T^{DH}_{ab}(x_a,x_b,A)=C A^{4/3} f_a/N(x_a) f_b/N(x_b), 
\label{tdh}
\end{equation}
where $C=0.35/(4r_0^2)$GeV$^2$, with $r_0=1.1-1.25$. At 
finite temperature, $T^{DH}_{ab}$ depends on the
density matrix at the given temperature. 

For a soft-hard subprocess, the contribution has the 
following general form\cite{GUO}:
\begin{equation}
\frac{d\sigma^{(SH)}}{dQ^2 dq_T^2 dy}
=\sum_{a,b,c}\, \int dx' f_{c/h}(x') \,
\Phi_{ac}^{SH}(x_a,x',A) \cdot \frac{1}{2x's} \cdot
\frac{12\pi\alpha_s^2}{x's+u-Q^2}\, \cdot 
\left(\frac{4\pi\alpha_{em}^2}{9Q^2} e_a^2 \right) \ ,  
\label{sh}
\end{equation}
with 
\begin{eqnarray}
\Phi_{ac}^{SH}(x_a,x',A) &=&
\left[ \frac{\partial^{2}}{\partial x_a^{2}}
\left( \frac{1}{x_a} T_{a}(x_a,A) H_{ac}^{SH}(x_a,x') \right)\right]
\cdot \frac{2q_{T}^{2}}{(x's+u-Q^2)^{2}} 
\nonumber \\
&\ & + \left[\frac{\partial}{\partial x_a}
\left( \frac{1}{x_a} T_{a}(x_a,A) H_{ac}^{SH}(x_a,x') \right) \right]
\cdot \frac{2(Q^2-u)}{x's(x's+u-Q^2)}  \ . 
\label{sh1}
\end{eqnarray}
In Eq.~(\ref{sh1}), $H_{ac}^{SH}$ is the calculated partonic part,
and $T_a(x_a,A)$ is the soft-hard matrix element.
For a cold nucleus,  LQS proposed the following model \cite{LQS}:
\begin{equation}
T_a(x_a, A)=\lambda^2 A^{4/3} f_{a/N}(x_a), 
\label{ta}
\end{equation}
where $\lambda^2 \sim 0.05-0.1$ GeV$^2$, and $\lambda$ is estimated
from the data on di-jet momentum imbalance \cite{LQS2}. 

Combining contributions from double-hard and soft-hard subprocesses, and 
using the models for $T^{DH}_{ab}$ and $T_a$, large 
nuclear enhancement for Drell-Yan production in hadron-nucleus 
collisions was obtained \cite{GUO}. A typical enhancement in large $q_T$ 
region is shown in Fig.~\ref{fig2}.

\section{SUPPRESSION IN SMALL $q_T$ REGION}

When $q_T\rightarrow 0$, $x_{1A} \sim x_{1B} \sim 0$, both soft-hard 
and double-hard contributions  
become divergent. But these divergences are canceled by the 
interference terms from Eq.~(\ref{sec5-e3}), because
\begin{equation}
M \rightarrow 
\left( \frac{F(x_{1A},x_{tot}-x_{1A})}{x_{1A}-x_{1B}} \right)_{soft-hard}
- 
\left( \frac{F(x_{1B},x_{tot}-x_{1B})}{x_{1A}-x_{1B}} \right)_{double-hard}
\rightarrow \quad \mbox{finite} \ ,
\end{equation}
as $q_T \rightarrow 0$. In addition, 
similar to the single-scattering, double-scattering
subprocesses develop the collinear and soft divergences 
when $q_T\rightarrow 0$.  Although all divergences are canceled after
including necessary virtual diagrams and proper collinear 
subtractions, there are large  
$\ell n(Q^2/q_T^2)$ for every power of $\alpha_s$.  A 
systematic resummation at the presence 
of double-scattering is necessary for a reliable prediction in 
small $q_T$ region \cite{GQS}.

Because of very {\it small} A-dependence in Drell-Yan $d\sigma /dQ^2$, 
and {\it large} nuclear enhancement for large $q_T$ region, one 
expects nuclear suppression in small $q_T$
region. This is consistent with above discussion on the role of 
quantum interference between the soft-hard and double-hard subprocesses.
It is known that nuclear shadowing in the parton distribution of
momentum fraction $x$ is due to quantum interference of multi-parton 
recombination in the longitudinal direction \cite{MQ}. 
If the same thought is applied to the
transverse direction, a nuclear suppression in 
small $q_T$ region is expected from quantum interference between 
soft-hard and double-hard subprocesses.  The
expected ratio of double-scattering
to single-scattering contribution in small $q_T$ region 
is sketched in Fig.~\ref{fig2} \cite{GQS}.

\end{document}